\documentclass{an}
\usepackage{graphicx}
\usepackage{times}
\usepackage{fancyhdr}
\begin{document}

\title{CCD-$\Delta a$ and $BVR$ photometry of NGC~7296\thanks{Based on observations at the Hvar Observatory, University of Zagreb and the Leopold-Figl Observatory for Astrophysics, University of Vienna.}}

\author{M. Netopil\inst{1} \and E. Paunzen\inst{1} \and H.M. Maitzen\inst{1} \and A. Claret\inst{2} \and K. Pavlovski\inst{3} \and E. Tamajo\inst{3}}
\institute{Institut f\"ur Astronomie der Universit\"at Wien, T\"urkenschanzstr. 17, A-1180 Wien,       
           Austria
\and	   Instituto de Astrof\'{\i}sica de Andaluc\'{\i}a
		   CSIC, Apartado 3004, 18080 Granada, Spain
\and       Department of Physics, University of Zagreb, Bijeni\u{c}ka 32, 10000 Zagreb, Croatia		   }

\date{Received; accepted; published online}

\abstract{The first CCD photometric investigation of the open cluster NGC~7296 up to now was performed within the narrow band $\Delta a$ photometric system, which enables us to detect peculiar objects. A deeper investigation of that cluster followed, using the standard $BVR$-Bessel filter set. The age and $E(B-V)$ was determined independently to log\,$t$\,=\,8.0\,$\pm$\,0.1 and 0.15\,$\pm$\,0.02 respectively by using $\Delta a$ and broadband photometry. In total five Be/Ae objects and two metal-weak stars showing significant negative $\Delta a$-values as well as one classical chemically peculiar star could be identified within that intermediate age open cluster.
}

\correspondence{netopil@astro.univie.ac.at}

\maketitle

\section{Introduction}

In continuation of previous work (Paunzen, Pintado \& Maitzen 2003 and references therein) and the search for chemically peculiar (CP) objects via CCD $\Delta a$-photometry the open cluster NGC~7296 has been investigated, the first photometric investigation of that object up to now. This photometric system measures the flux depression at 5200\,\AA, a typical feature for CP and related objects (Kupka et al. 2004). The prerequisite for investigation larger samples of CP stars (including the
generally fainter open cluster members) is unambiguous detection. The $\Delta a$-photometry is therefore an economic way to detect them also in more distant open clusters and will be used for a object selection of further spectroscopic surveys, also a statistical analysis of the detected peculiar stars is already in preparation. For calibration uses and a deeper study we have obtained additional $BVR$ observations of that cluster, which is also the start of a project dedicated to ``forgotten'' open clusters to fill the white spots on the galactic map. With the help of the corresponding isochrones for the $\Delta a$ system (Claret, Paunzen \& Maitzen 2003; Claret 2004) the age and the distance were determined to $100_{-21}^{+26}$\,Myr and 2900\,$\pm$\,350\,pc respectively. A verification was also performed with independent isochrones for the broadband measurements (Girardi et al. 2000) yielding excellent agreement. We have detected eight possible peculiar stars in the field of NGC~7296 which are probable members according to their position within the Hertzsprung-Russell-Diagram (HRD).       

\section{Observations and reduction}
\subsection{$\Delta a$ photometry}
The $\Delta a$ photometric system itself was defined by Maitzen (1976). It measures the broad band absorption feature at 5200\,\AA$ $ sampling the depth of this flux depression by comparing the flux at the center (5220\,\AA, $g_{2}$), with the adjacent regions (5000\,\AA, $g_{1}$ and 5500\,\AA, $y$). The respective index was introduced as 

$a = g_{2} - (g_{1} + y) / 2$.

Since this quantity is slightly dependent on temperature (increasing towards lower temperatures), the intrinsic peculiarity index had to be defined as

$\Delta a = a-a_{0}[(b-y);(B-V);(g_{1}-y)]$

i.e. the difference between the individual $a$-value and the $a$-value of non-peculiar stars of the same colour (the locus of the $a_{0}$-values has been called normality line). It was shown (e.g. Vogt et al. 1998) that virtually all peculiar stars with magnetic fields (CP2 stars) have positive $\Delta a$-values up to 0.075\,mag whereas Be and $\lambda$\,Bootis stars exhibit significant negative ones. Note that $(g_{1}-y)$ shows an excellent correlation with $(b-y)$ or $(B-V)$ and can be used as an index for the effective temperature.

The observations were performed on 2004 August 8 at the Hvar Observatory, University of Zagreb, using the 1-m Austro-Croatian Telescope (ACT). The telescope is equipped with a SiTe003AB CCD chip of 1024x1024\,px, allowing a field-of-view of 5.7'. The exposure times were limited to 150 seconds, enough to cover stars down to V$\approx$15\,mag at the needed accuracy. The basic CCD reductions 
(bias-subtraction, dark-correction, flat-fielding) were carried out within standard IRAF V2.12.2 routines. For all frames we have applied a point-spread-function-fitting within the IRAF task 
DAOPHOT (Stetson 1987). Photometry of each frame was performed separately and the measurements were then averaged and weighted by their individual photometric error. 

As diagnostic tool the $a$ versus $(g_{1}-y)$ diagram is used. Assuming that all stars exhibit the same interstellar reddening, peculiar objects deviate from the normality line more than 3$\sigma$. An $a$ versus $(b-y)_{0}$ or $(B-V)_{0}$ diagram should then be able to sort out the true peculiar objects.

\subsection{$BVR$ photometry}
Additional CCD-$BVR$ observations were obtained on 2004 September 9 at the 1.52-m telescope of the 
Leopold-Figl Observatory for Astrophysics (FOA) of the University of Vienna, located at the Mittersch\"opfl (Austria) using the multimode instrument OEFOSC and a SITe SI502AB CCD as detector, which provides a field-of-view of 5.75'.

The basic CCD reductions and point-spread-function-fitting are carried out in the same manner like for the $\Delta a$ photometry. In addition to the programme cluster we have observed a standard field of NGC~7654 by Stetson (2000), used for the transformation to standard magnitudes. Since the programme cluster as well as the standard field were obtained at the zenith with an airmass between 1.00\,-\,1.01 directly one after another, no atmospheric extinction correction was necessary. After applying the aperture correction using fourteen bright isolated stars, the transformation between standard and instrumental magnitudes was performed using the coefficients given in Table \ref{regression}. Since no clear colour dependency was found, it was left out instead of adding additional errors to the photometric measurements. The typical errors for the individual $BVR$-photometry are up to 0.05\,mag and are also given in the tables provided electronically.

The observing log with the number of frames in each filter is listed in Table \ref{Obslog}. Exposure times were varied from 10 to 180 seconds, depending on the used filter, to cover a larger magnitude range. We have to note, that the $R$ measurements for BD+51\,3383 have to be used with caution, since this star was nearly saturated and therefore out of the linearity range, also on frames with the shortest exposure time.

\begin{table}
\caption{The number of frames taken in each filter system.}
\label{Obslog}
\begin{tabular}{cccccl}\hline
Site & Date & \# & \# & \# & Filter\\
\hline
ACT & 2004 Aug. 8 & 6 & 7  & 6 & $g_{1}/g_{2}/y$\\
FOA & 2004 Sep. 9 & 8 & 12 & 13 & $B/V/R$\\ 
\hline
\end{tabular}
\end{table}

\begin{figure}
\resizebox{\hsize}{!}
{\includegraphics[]{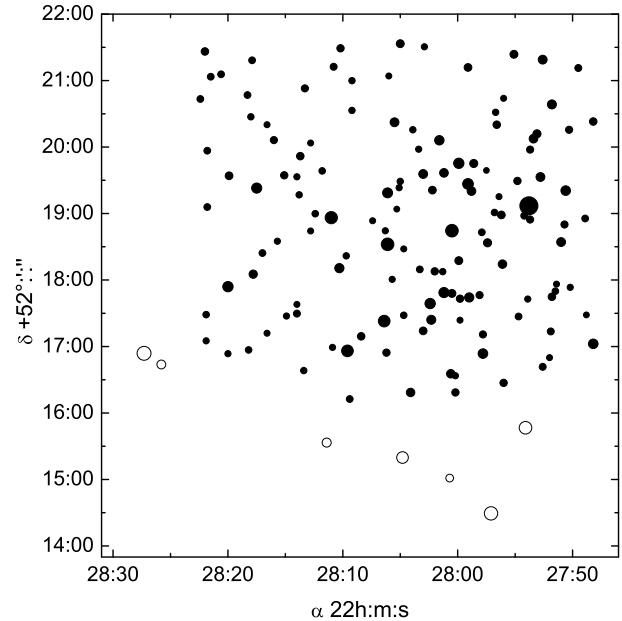}}
\caption{The observed field of NGC~7296. Stars measured only within the $\Delta a$ system are marked with open circles.}
\label{figure1}
\end{figure}

\begin{table}
\caption{The regression coefficients for the transformations from instrumental $bvr$ to standard $BVR$ magnitudes (upper part) and the calibration of the $\Delta a$ system as well as the coefficients for  the normality line (lower part). The errors in the final digits of the corresponding quantity are given in parenthesis.}
\label{regression}
\begin{tabular}{clc}\hline
 & regression coefficients & N \\
\hline
$B$     & $= +1.90(9) + 1.012(12)\cdot b$ & 24\\
$V$     & $= +2.55(8) + 0.999(13)\cdot v$ & 24\\
$R$     & $= +2.93(7) + 1.004(16)\cdot r$ & 24\\
\hline
$V$     & $= -0.19(7) + 1.004(5)\cdot y$ & 34\\
$(B-V)$ & $= +0.06(1) + 2.116(45)\cdot(g_{1}-y)$ & 34\\
$(V-R)$ & $= -0.03(1) + 1.523(44)\cdot(g_{1}-y)$ & 34\\
$a_{0}$ & $= +0.267(2) + 0.285(16)\cdot(g_{1}-y)$ & 21\\
\hline
\end{tabular}
\end{table}

\section{Results}
In the following we will discuss the results for NGC~7296 in more detail. This is a moderately rich cluster with a Trumpler type classification of II\,2\,p (Lyng{\aa} 1987) and a well defined sharp main sequence, showing no evidence for differential reddening. Since most of the observed 140 stars are located on or near the main sequence, a clue for their membership as a first approximation, the richness class has to be set at least to 'm'. The majority of the stars observed within the $\Delta a$ system were also investigated in the broadband. For seven stars only $\Delta a$ measurements are available. These stars were calibrated to the standard $BVR$ system using the correlations given in Table \ref{regression} and are marked in all diagrams with different symbols. For these stars the pixel coordinates were transformed to positions on the CCD mounted at the FOA, resulting in a larger error due to the slightly tilted camera at ACT (Figure \ref{figure1}).

We have identified eight objects deviating more than 20\,mmag from the normality line, which was derived with probable members only. The probable nonmembers were identified by their location within the HRD. The slope is dominated on the first sight by the giant BD+51\,3383. Calculating it without that star, the zero-point remains identical and the slope decreases only by 1\,\%, causing only negligible differences in the pecularity index of the deviating stars, so we included that giant resulting in the regression coefficients given in Table \ref{regression}.

\begin{figure}
\resizebox{\hsize}{!}
{\includegraphics[]{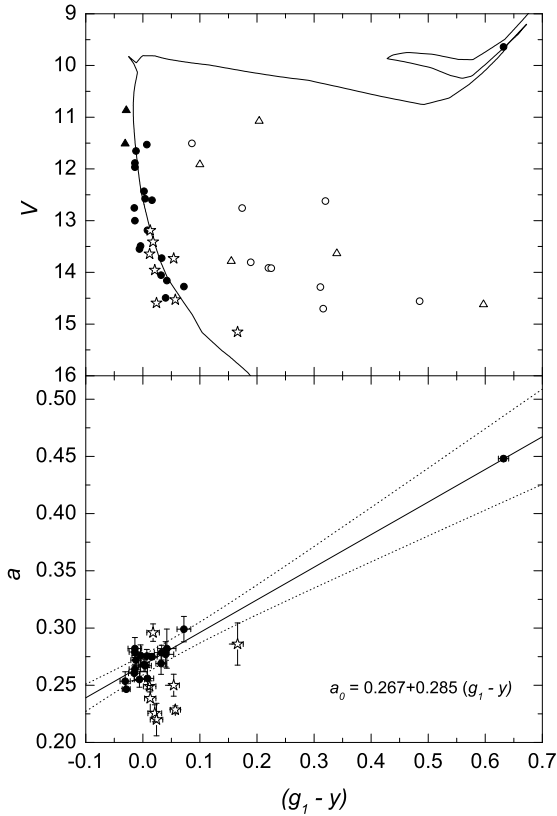}}
\caption{Observed V versus $(g_{1}-y)$ and $a$ versus $(g_{1}-y)$ diagrams for NGC 7296. The solid line is the normality line whereas the dotted lines are the confidence intervals corresponding to 99.9\%. The error bars for each individual object are the mean errors. The detected peculiar objects are marked with open asterisks, stars with calibrated V magnitudes are indicated by triangles. Only members (filled symbols) excluding peculiar stars have been used to derive the normality line. The isochrone was taken from Claret et al. (2003) which is based on the $\Delta a$ photometric system only, resulting in log\,$t$\,=\,8.0\,$\pm$\,0.1, $m_{V}-M_{V}$\,=\,12.8\,$\pm$\,0.4, and $E(B-V)$\,=\,0.07\,$\pm$\,0.06.}
\label{figure2}
\end{figure}

\begin{figure}
\resizebox{\hsize}{!}
{\includegraphics[]{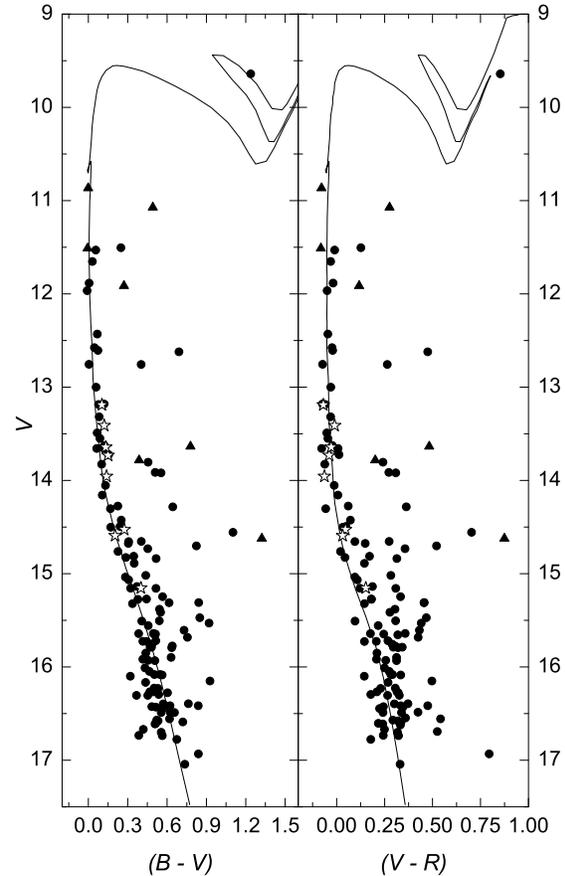}}
\caption{Observed $V$ versus $(B-V)$ and $V$ versus $(V-R)$ diagrams for NGC~7296. Stars with calibrated magnitudes are indicated by triangles, the apparent peculiar objects with open asterisks. The isochrones were taken from Girardi et al. (2000), resulting in a best fit with the parameters log\,$t$\,=\,8.0\,$\pm$\,0.1, $m_{V}-M_{V}$\,=\,12.8\,$\pm$\,0.2, and $E(B-V)$\,=\,0.15\,$\pm$\,0.02.}
\label{figure3}
\end{figure}

Five late B and early A-type stars (\#19,\,22,\,23,\,24,\,30 according to our numbering system) show significant negative $\Delta a$-values and seem to be Be/Ae objects. The stars \#20 and 50, also showing a negative $\Delta a$ index, are good candidates for metal-weak objects, because they are too cool for being B-type or early A-type stars, whereas the position of star \#50 in the colour-magnitude diagram (CMD), especially seen in Figure \ref{figure2}, challenges its membership status. One classical CP star (\#14) was also detected with a moderate $\Delta a$ index of 23\,mmag. The results for all peculiar stars are given in Table \ref{ResultsCP}.

\begin{table}
\caption{Summary of results for the detected peculiar objects. The star numbers are according to our numbering system. The type of peculiarity is given in the last column, whereas mw stands for metal-weak stars.}
\label{ResultsCP}
\begin{tabular}{lccccc}\hline
Star & $\Delta a$ [mmag] & V & $(B-V)$ & $(g_{1}-y)$ & pec.\\
\hline
14 & $+$23 & 13.412 & $+$0.121 & $+$0.013 & CP2\\
19 & $-$48 & 13.958 & $+$0.139 & $+$0.016 & Be/Ae\\
20 & $-$55 & 14.528 & $+$0.274 & $+$0.052 & mw\\
22 & $-$33 & 13.191 & $+$0.103 & $+$0.008 & Be/Ae\\
23 & $-$33 & 13.732 & $+$0.150 & $+$0.049 & Be/Ae\\
24 & $-$22 & 13.644 & $+$0.133 & $+$0.007 & Be/Ae\\
30 & $-$54 & 14.593 & $+$0.205 & $+$0.019 & Be/Ae\\
50 & $-$29 & 15.154 & $+$0.402 & $+$0.161 & mw\\
\hline
\end{tabular}
\end{table}

Our $BVR$ photometry compared to three stars also observed by Tycho (Hog et al. 2000) shows a partial agreement. The measurements by Tycho were corrected with the terms given by Bessel (2000) resulting in deviations to our results which are given in Table \ref{tycho}. Due to the small sample of comparison stars and since the deviations are either within the errors or very large alternating within $V$ and $(B-V)$, no clear conclusion about possible zero points can be given. We have to note, that except BD+51\,3383 all other stars were observed within the $\Delta a$ system only and are transformed to $V$ or $(B-V)$ respectively according to the correlations given in Table \ref{regression}. However that transformation can not be made responsible for some large deviations up to 0.5\,mag. 

\begin{table}
\caption{The differences between our $BV$-photometry and the one by Tycho ($(B-V)-(B_{T}-V_{T})$).}
\label{tycho}
\begin{tabular}{llcc}\hline
Star & & $\Delta (B-V)$\,[mag] & $\Delta V$\,[mag]\\
\hline
122 & BD+51\,3383 & $-$0.26 & $-$0.13 \\
134 & GSC 03619-01339 & $+$0.03 & $-$0.04 \\
136 & GSC 03619-01557 & $-$0.07 & $-$0.49 \\
\hline
\end{tabular}
\end{table}

\begin{table}[h]
\caption{The parameter results for NGC~7296. The errors in the final digits of the corresponding quantity are given in parenthesis.}
\label{parameters}
\begin{tabular}{lc}
\hline
NGC~7296 (C2226+520)\\
\hline
$l/b$ & 101.9/$-$4.6 \\
$E(B-V)$ & 0.15(2) \\
$E(V-R)$ & 0.00(2) \\
$m_{V}-M_{V}$ & 12.8(2) \\
d [kpc] & 2.93(35) \\
log\,$t$ & 8.0(1) \\
\hline
\end{tabular}
\end{table}

The age of NGC~7296 was determined to log\,$t$\,=\,8.0\,$\pm$\,0.1 using the corresponding isochrones for the $\Delta a$ system given in Claret et al. (2003) which are based on the stellar models published in Claret (2004). To obtain an independent comparison, we have used the isochrones by Girardi et al. (2000) for our broadband photometry resulting in the same age for the best fit to the CMD, also if neglecting the evolved star BD+51\,3383, which membership seems to be sure using two independent photometric systems. Furthermore the apparent distance module $m_{V}-M_{V}$\,=\,12.8\,$\pm$\,0.2 and the colour excesses $E(B-V)$\,=\,0.15\,$\pm$\,0.02 as well as $E(V-R)$\,=\,0.00\,$\pm$\,0.02 were defined, yielding a distance of 2930\,$\pm$\,350\,pc using the standard absorption ratio $R_{V}$=3.1. The distance module and reddening values are based more precisely on the broadband photometry, since the $\Delta a$-CMD with its limiting magnitude gives a lot of possibilities for parameter combinations. A summary of all results and their errors can be found in Table \ref{parameters}.

The relative high number of Be stars at that age found in this cluster is 
comparable to IC~4725 and NGC~3114 which contain 9 and 4 Be stars respectively (Lyng{\aa} 1987). According to Abt (1979) the frequency of Be stars in open clusters shows no obvious variation with age, what is in contradiction to Mermilliod (1982) who found a decrease towards older ages. Since the fraction of Be stars is increasing to lower metallicities (Maeder, Grebel \& Mermilliod 1999), the result for NGC~7296 can be explained with a different environment, but the evolutionary status of Be stars is still a unsolved question (Fabregat \& Torrej\'on 2000).

\section{Conclusion}
Using the first photometric investigation of NGC~7296 within the $\Delta a$ as well as the $BVR$ photometric system, the age was determined by means of three colour-magnitude diagrams and the help of two different isochrone calculations to $100_{-21}^{+26}$\,Myr. Despite some divergent deviations of the $BV$-photometry compared to the one by Tycho (Table \ref{tycho}), the obtained apparent distance module of 12.8\,$\pm$\,0.2 leads to a distance of 2.93\,$\pm$\,0.35\,kpc using $E(B-V)$\,=\,0.15\,$\pm$\,0.02 and the standard absorption ratio $R_{V}$\,=\,3.1. We have found eight peculiar objects deviating more than 20\,mmag from the normality line, therefrom five stars are indicating to be Be/Ae objects. Furthermore two metal-weak as well as one classical chemical peculiar star were detected. However the membership status for these stars can only be determined by means of their location within the colour-magnitude diagrams. 

The tables with all the data of the observed stars are available in electronic form at the CDS via anonymous ftp to cdsarc.u-strasbg.fr (130.79.125.5), http://cdsweb.u-strasbg.fr/Abstract.html or upon request from the first author. These tables include the coordinates ($X/Y$ within our frames, $\alpha$/$\delta$), the observed $(g_{1}-y)$ and $a$ values, $\Delta a$-values derived from the normality line of $(g_{1}-y)$ (exclusive nonmembers), $V$ magnitudes, the $(B-V)$ and $(V-R)$ colours as well as the corresponding errors.

\acknowledgements
This research was performed within the project {\sl 10/2004} of the WTZ (Wissenschaftlich Technische Zusammenarbeit Kroatien) as well as {\sl P17580} and {\sl P17920} of the Austrian Fonds zur F{\"o}rderung der wissen\-schaft\-lichen Forschung (FwF) and benefited also from the financial contributions of the City of Vienna (Hochschuljubil{\"a}umsstiftung project: $\Delta a$ Photometrie in der Milchstrasse und den Magellanschen Wolken, H-1123/2002). One of us (MN) acknowledges the support by a ``Forschungsstipendium'' of the University of Vienna. Use was made of the SIMBAD database, operated at CDS, Strasbourg, France and the WEBDA database, operated at the Institute of Astronomy of the University of Lausanne. This research has made use of NASA's Astrophysics Data System.


\begin{thebibliography}{}
\bibitem{} Abt, H.A..: 1979, ApJ 230, 485
\bibitem{} Bessel, M.S.: 2000, PASP 112, 961
\bibitem{} Claret, A.: 2004, A\&A, 424, 919
\bibitem{} Claret, A., Paunzen, E., Maitzen, H.M.: 2003, A\&A, 412, 91
\bibitem{} Fabregat, J., Torrej\'on, J.M.: 2000, A\&A 357, 451
\bibitem{} Girardi, L., Bressan, A., Bertelli, G., Chiosi, C.: 2000, A\&AS 141, 371
\bibitem{} Hog, E., Fabricius, C., Makarov, V.V. et al.: 2000, A\&A 355, L27 
\bibitem{} Kupka, F., Paunzen, E., Iliev, I. Kh., Maitzen, H.M.: 2004, MNRAS 352, 863
\bibitem{} Lyng{\aa}, G.: 1987, Catalogue of Open Cluster Data, 5th edition, CDS, Strasbourg
\bibitem{} Mermilliod, J.C.: 1982, A\&A 109, 48
\bibitem{} Maitzen, H.M.: 1976, A\&A 51, 223
\bibitem{} Maeder, A., Grebel, E.K., Mermilliod, J.C.: 1999, A\&A 51, 223
\bibitem{} Paunzen, E., Pintado, O. I., Maitzen, H.M.: 2003, A\&A 412, 721
\bibitem{} Stetson, P.B.: 1987, PASP 99, 191
\bibitem{} Stetson, P.B.: 2000, PASP 112, 925
\bibitem{} Vogt, N., Kerschbaum, F., Maitzen, H.M., Faundez-Abans, M.: 1998, A\&AS 130, 455
\end{thebibliography}
\end{document}